\documentclass[pre,preprint,
showpacs]{revtex4}

\usepackage{graphicx}%
\usepackage{dcolumn}
\usepackage{amsmath}

\begin{document}
\title[]{High Temperature Electron Localization in Dense He 
Gas}

\author{A.F.Borghesani 
and M.Santini}
\affiliation{Istituto Nazionale per la Fisica della Materia\\
 Dipartimento di Fisica ``G.Galilei'', 
Universit\`a di Padova \\via F. Marzolo 8, I-35131 Padova, Italy}
\begin{abstract}
We report new accurate mesasurements of the mobility of excess electrons in high
density Helium gas in extended ranges of temperature
$[(26\leq T\leq 77 )\, {K} ] $ and density $ [ (0.05\leq N\leq 12.0) $
$ {atoms} \cdot {nm}^{-3}] $ to ascertain the effect of temperature on
the formation and dynamics of localized electron states.  The main
result of the experiment is that the formation of localized states
essentially depends on the relative balance of fluid dilation energy, 
repulsive electron--atom interaction energy, and thermal energy. As a 
consequence, the onset of localization depends on the medium disorder 
through gas temperature and density.
It appears that the transition from
delocalized to localized states shifts to larger densities as the
temperature is increased.  This behavior can be understood in terms of
a simple model of electron self--trapping in a spherically symmetric
square well.
\end{abstract}

\maketitle

\section{Introduction }
\label{sec:intro}
The transport properties of excess electrons in dense noble gases and
liquids give useful information on the electron states in a disordered
medium and on the relationship between the electron--atom interaction
and the properties of the fluid.  The electron behavior depends on the 
strength of its coupling with the gas atoms and on the response function
of the gas itself.  Therefore, different transport mechanisms and
regimes can be obtained according to the nature of the electron--atom
interaction (repulsive or attractive), to the thermodynamic conditions
of the gas, either close to or removed from its critical point, and to
the amount of disorder inherent to the fluid \cite{hern91}.

Typically, at low density and high temperature, electrons are
quasifree.  Their wavefunction is pretty delocalized and the resulting
mobility is large.  They scatter elastically off the atoms of noble
gases in a series of binary collisions and the scattering process is
basically determined by the interaction potential through the
electron--atom scattering cross section.  The mobility can be
predicted accurately by the classical kinetic theory \cite{hux74}.

At higher densities, and, possibly, at lower temperatures, electrons
may either remain quasifree with large mobility (as in the case of
Argon), or they can give origin to a new type of state that is
spatially localized inside a dilation of the fluid.  In this case the
mobility is very low because the complex electron plus fluid dilation
moves as an unique, massive entity.  This, for instance, happens in He
and Ne.  The main difference between the two cases is that in the
former the electron--atom interaction is attractive (Ar) and in the
latter is repulsive (He and Ne) \cite{borg941,borg942}.

The simplest model to describe the behavior of electrons in a dense,
disordered medium is the hard--sphere gas and a practical realization
of this system is represented by He.  In He the electron--atom
interaction is pretty well described by a hard--core potential and the
scattering cross section is fairly large and energy independent.  It
is well known that the charge transport proceeds via bubble formation
in liquid He at low temperature \cite{levi,harri,sch80,ja}. In gaseous 
He at low temperature
the mobility shows a drop of several orders of magnitude when the
density is increased from low to medium values.  This drop has been
intepreted in terms of a continuous transition from a transport regime
where the excess electrons are quasifree to a region where they are
localized.  There is still controversy about the nature of the
localized states, whether they are localized in bubbles as in the case
of the liquid or whether they are localized in the Anderson sense 
\cite{poli}.  In
this case the electron wavefunction decays exponentially with the
distance owing to multiple scattering effects induced by the disorder
of the medium \cite{om}.

Owing to these considerations, it is interesting to investigate the
localization transition at higher temperatures. 
Therefore, we have measured the mobility of excess
electrons in dense He gas at temperatures $26<T<77\, K.$ By assuming
that electrons are localized in dilations of the gas, a simple quantum
mechanical model provides a good semiquantitative description of the
observed behavior of the mobility.

\section{Experimental Details}\label{sec:exptl}
\begin{figure}[htbp]
    \centering
    \includegraphics[scale=0.7]{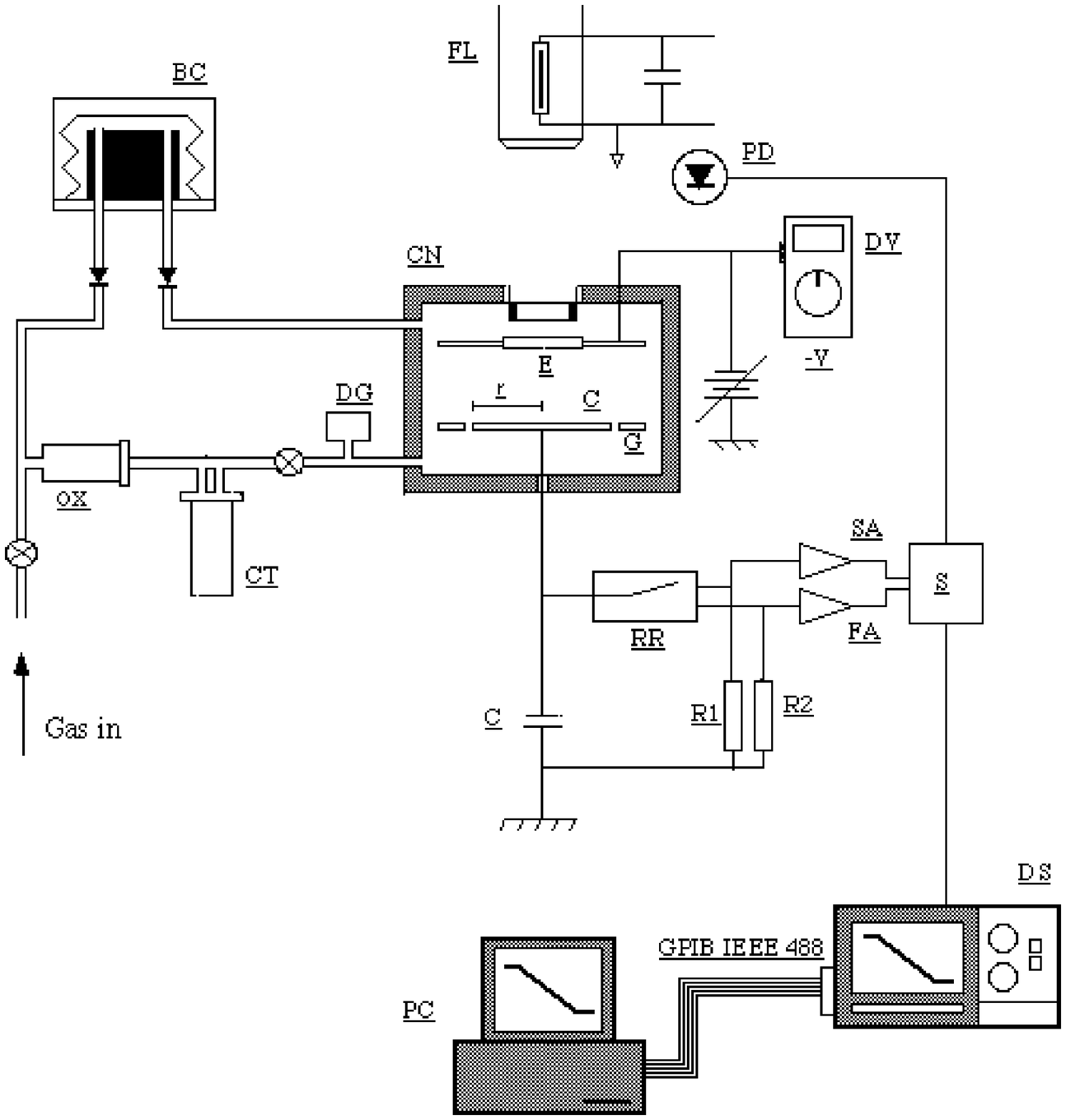}
    \caption{\small Schematics of the experimental apparatus. See the text for a 
description.}
    \label{fig:fig1}
\end{figure}

The mobility measurements have been carried out by using a swarm
technique in a Pulsed Townsend Photoinjection apparatus we have been
exploiting for a long time for electron and ion mobility measurements
\cite{borg88,borg90,borg93}. A schematics of the apparatus is shown in 
Figure \ref{fig:fig1}.
Briefly, 
a high-pressure cell (CN), that can withstand pressures up to 10 MPa, 
is mounted on the cold head of a cryocooler inside a triple-shield thermostat. 
The cell is operated between 25 and 330 K. Temperature is stabilized 
within 0.01 K. 

A parallel-plate capacitor, consisting of an emitter (E) and a collector (C), 
is contained in the high-pressure cell and is energized by the high-voltage 
generator -V. A digital voltmeter (DV) reads the voltage. The distance 
between the two plates delimits the drift space. An electron swarm is produced by 
irradiating the gold-coated quartz window placed in the emitter with the VUV 
light pulse of a Xe flashlamp (FL). The amount of produced charge ranges between
4 and 400 fC, depending on the gas pressure and on the applied electrical field 
strength. Under its action, the charges drift towards the anode inducing a 
current in the external circuit. The current is integrated by the analog 
circuit RC in order to improve the signal-to-noise ratio. Two different 
operational amplifiers (SA and FA) are used depending on the duration of 
the signal. This is recorded by a high-speed digital transient analyzer 
(DS) and is fetched by a personal computer for the analysis of the waveform.

Ultra-high purity He gas with an impurity content, essentially Oxygen, 
of some p.p.m is used. The impurity content is reduced to a few 
p.p.b. by circulating the gas in a recirculation loop driven 
by a home-made bellow circulator (BC) that forces the gas to be purified 
through an Oxisorb cartridge (OX) and a LN2-cooled active-charcoal trap (CT).

The induced signal waveform of the electron drifting at constant 
speed is a straight line, and the drift time is easily 
determined by the analysis of the waveform. The overall accuracy of 
the mobility measurements is $ \vert\Delta \mu / \mu\vert\approx 5 
\,\% .$

\section{Experimental Results}\label{sec:res}
\begin{figure}[htbp]
    \centering
    \includegraphics[scale=0.6]{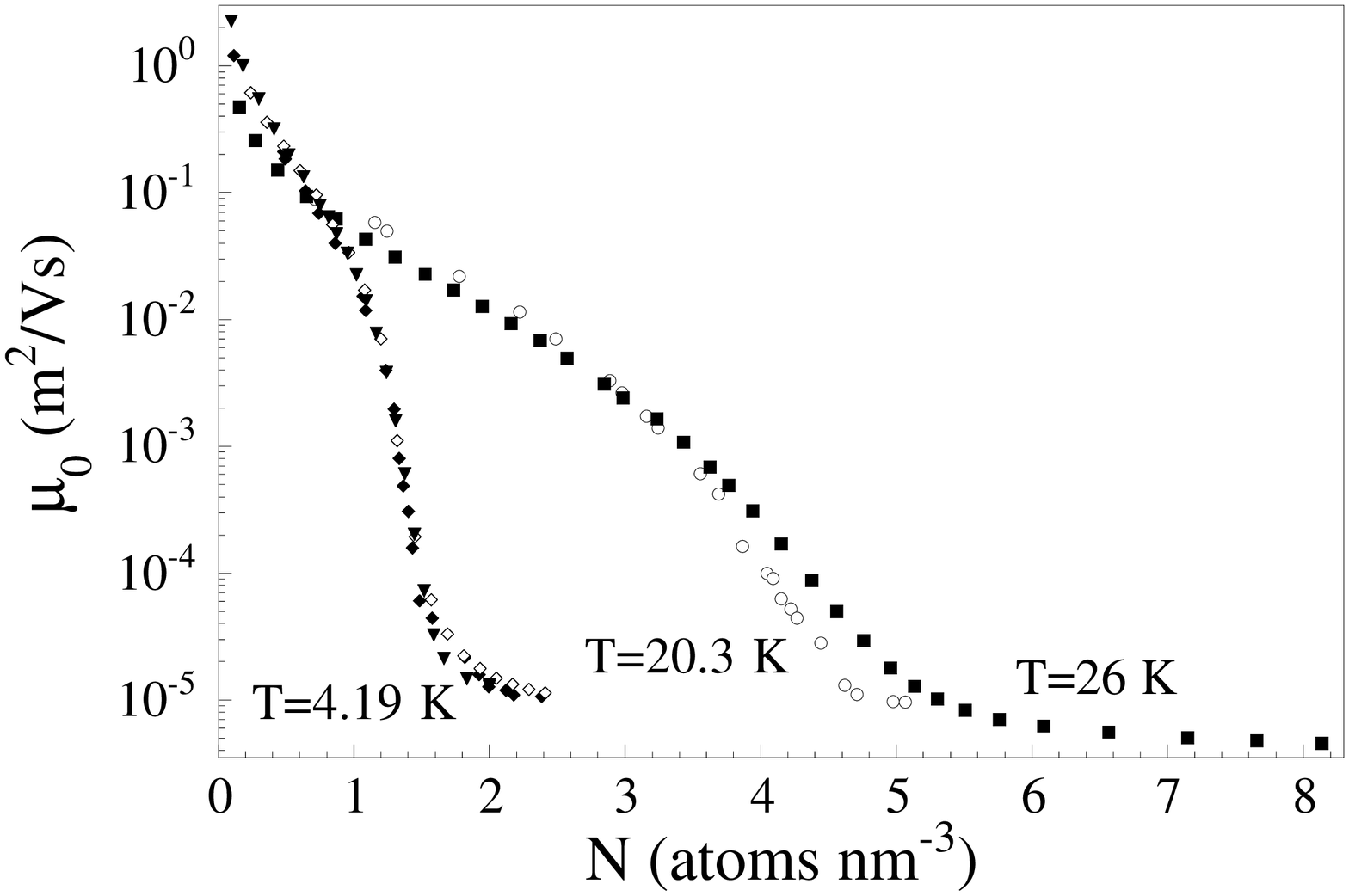}
    \caption{\small Experimental zero--field mobility $ \mu_{0}$ 
as a function of the 
gas density. $   T=26\, K:$
present work; $  T<26 K: $ literature 
data at $T\approx 4\, K$ \cite{levi,harri,sch80} and at $T\approx 20\, 
K$ \cite{ja}.}
\label{fig:fig2}
\end{figure}
\noindent In Figure \ref{fig:fig2}
we show the observed zero--field mobility $\mu_{0}$ in He 
at $T\approx 26\, K.$ The present data are compared with literature data for $T=4.2 \, K$ 
\cite{levi,harri,sch80} 
and for $T=20.3\, K$ \cite{ja}. At $T=26\, K$ $\mu_{0}$ exhibits the same 
qualitative behavior observed earlier at much lower temperatures. As the gas 
density increases, $\mu_{0}$ decreases by nearly 5 orders of magnitude.
The continuous transition from the low--density high--mobility region 
is interpreted as the progressive depletion of extended or delocalized 
states and the consequent formation of localized states 
\cite{poli,iaku1}. These are 
assumed to consist of an electron trapped into a cavity in the fluid. 
This cavity is referred to as an electronic bubble. A similar 
physical process 
has been observed also in liquid \cite{bruschi} and gaseous Neon 
\cite{borg90}. In gaseous Neon, the $\mu_{0}$ data resemble closely 
those shown in Figure \ref{fig:fig2} and the interpretation of the 
Neon data, as due to electron localization in cavities, has been 
confirmed by quantum--mechanical Molecular Dynamics calculations 
\cite{ancilotto}.

The dynamics of 
the localization process, though not investigated experimentally, is 
quite clear \cite{rose1,rose2,sakai}. However, even if the
localization process were of the Anderson type \cite{poli}, i.e., 
electrons with energy below the mobility edge trapped as a 
consequence of the self--interference of their wavefunction because of 
the medium disorder, nonetheless electrons wind up by forming electron 
bubbles because of the repulsive electron--medium interaction and 
medium compliance. The existence of such bubbles has been also 
confirmed experimentally by infrared absorption spectra in liquid He 
\cite{adams1, adams2}. 

Once all of 
the electron states are localized, the resulting $\mu_{0}$ is not zero 
because the gas is compliant enough to allow the large complex 
structure made of an electron plus the associated bubble to  
diffuse slowly \cite{hern91} and drift under the action of an external 
electric field..

The main difference between the present data and those at lower temperatures is that the 
transition to low mobility states is shifted to larger values of the 
density. At $T=4.2\, K$ the transition can be considered complete at a 
density $N\approx 2\, atoms \cdot nm^{-3}.$ At $T\approx 20.3 \, K$ 
the final state is reached for $N\approx 4.8 \, atoms \cdot nm^{-3},$ 
while at $T=26 \, K$ in our experiment this density has moved to 
$N\approx 6.2\, atoms\cdot nm^{-3}.$ This is even more evident at 
higher temperatures.

It is evident that the formation 
of localized states is not related to the presence of a nearby 
critical point (the critical temperature of He is $T_{c}\approx 5.2\, 
K ).$ It rather seems related to the competition between the 
thermal energy of electrons and the free energy of localization. 
Therefore, it appears reasonable that the localization transition 
shifts at larger densities for higher temperatures in order to 
achieve increasingly larger free energies.

The localization transition can be noticed also by observing the 
electric field dependence of the mobility. 
\begin{figure}[htbp]
    \centering
    \includegraphics[scale=0.6]{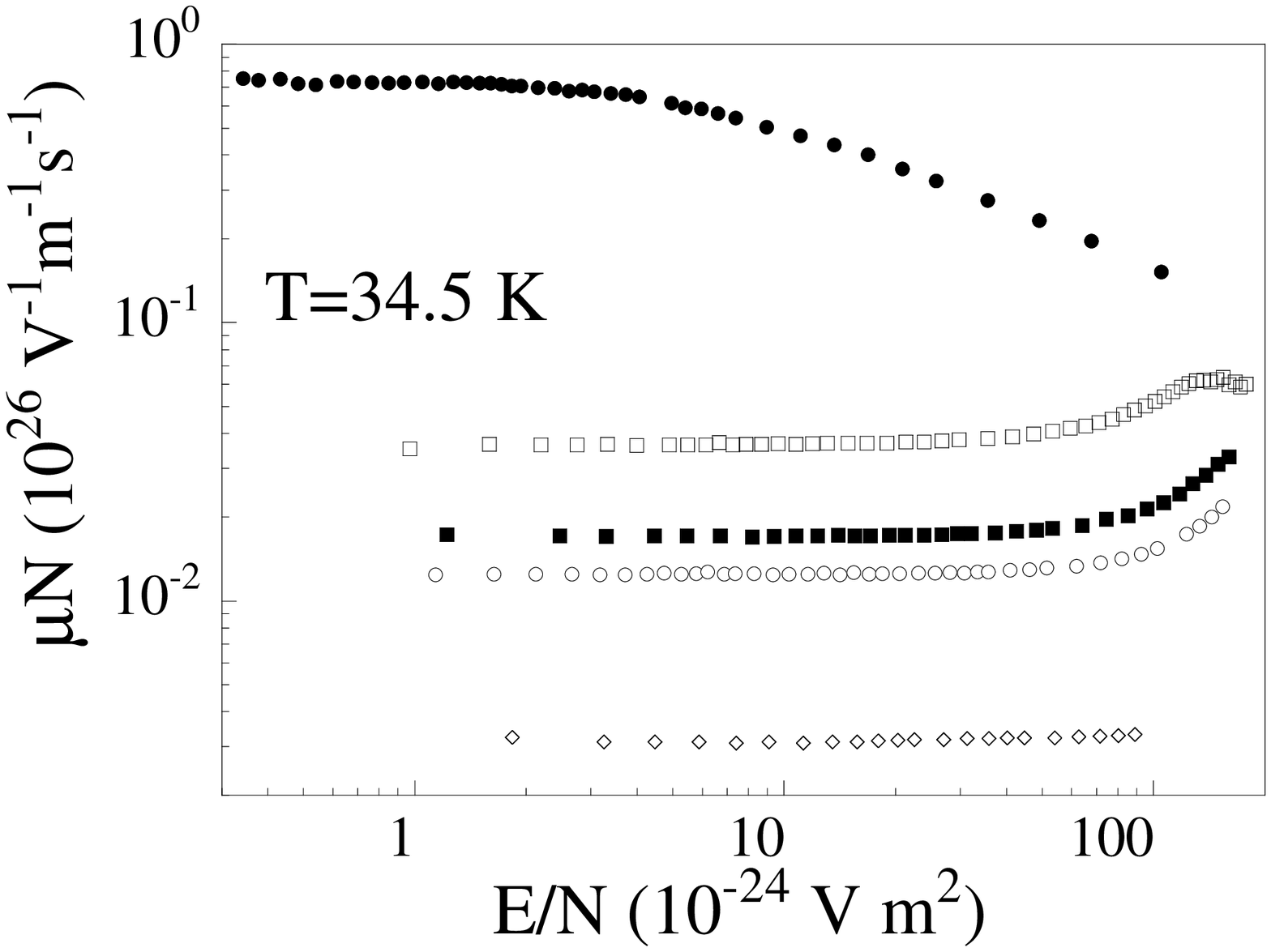}
    \caption{\small Density--normalized mobility $   \mu N$ as a function of 
the reduced electric field $   E/N$ at $
T=34.5\, K$ for several densities: $   N= 0.154,\, 4.174,\, 
4.658,$ $   4.833,\, 5.558\, atoms \cdot nm^{-3}$ (from top).}
    \label{fig:fig3}
\end{figure}
\noindent In Figure \ref{fig:fig3} we plot the 
density--normalized mobility $\mu N$ as a function of the reduced 
field $E/N$ at $T=34.5\, K$ for several densities. The behavior at 
different temperatures is similar to this one. 

At small $N$ and low 
$E/N,$ electrons are in near thermal equilibrium with the gas atoms 
and $\mu N$ is constant. As $E/N$ increases, $\mu N$ decreases, 
eventually reaching the $(E/N)^{-1/2}$ dependence expected on the 
basis of the classical kinetic theory because the scattering rate 
increases with the electron kinetic energy \cite{hux74}.

At high $N,$ $\mu N$ is very low and practically independent of 
E/$N,$ at least for the highest electric fields of the present experiment 
(up to $\approx 7\, kV/cm $).
At such densities, almost 
all of the 
electrons are localized in bubbles. Even the highest electric 
field reached in the experiment is not large enough to heat up such massive 
objects. The electronic bubbles therefore remain in equilibrium with the 
gas atoms.

At intermediate values of $N,$ the behavior of $\mu N$ is quite 
complicated. At small $E/N,$ $\mu N$ is constant, while at larger $E/N,$ $\mu 
N$ reaches a maximum and finally, at even larger $E/N,$ it meets  
 the classical $(E/N)^{-1/2}$ behavior. The same superlinear behavior 
 of the drift velocity of electrons in dense He gas was observed also 
 at very low 
 temperatures \cite{sch80}.

 This behavior can be interpreted in terms of 
the formation, at large $N,$ of electron states 
which are self--trapped in partially filled 
bubbles. These are very massive and have low mobility. By increasing 
the electric field strength bubbles may be either destroyed or their 
formation may be inhibited, 
so that electrons are again free and very mobile. The same  
behavior of $\mu N$ as a function of $E/N$ has 
been observed also in Neon gas and the same 
interpretation of the data has proven successful \cite{borg90}. 
\begin{figure}[htbp]
    \centering
    \includegraphics[scale=0.6]{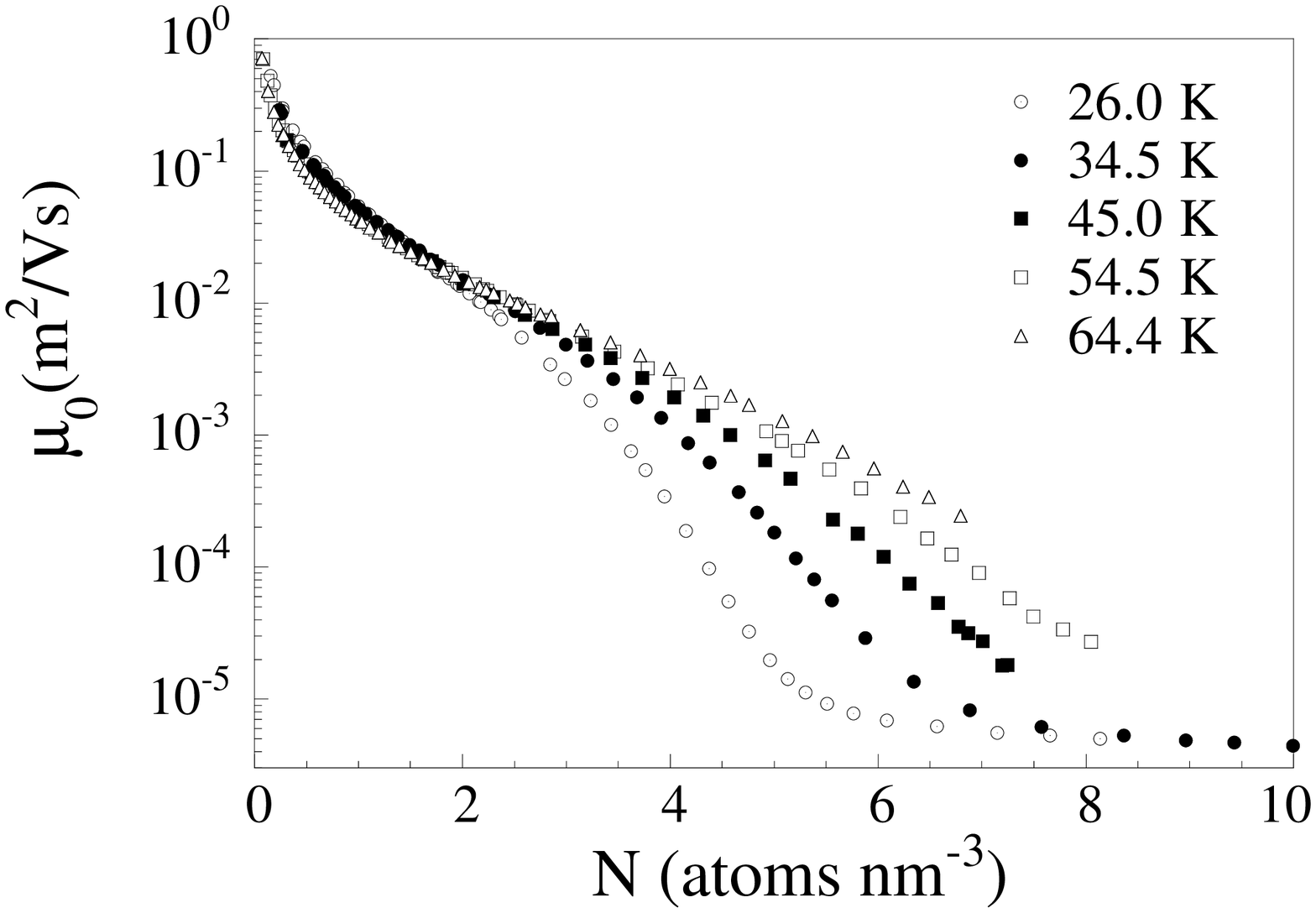}
    \caption{\small Zero--field mobility $  \mu_{0}$ 
as a function of $   N$ for $
  T= 26,\, 34.5,\, 45.0$ $   54.5,\, 64.4 \,K.$}
    \label{fig:fig4}
\end{figure}
In Figure \ref{fig:fig4},
 the zero--field value $\mu_{0}$ of the 
mobility $\mu $ is shown as a function of the density $N$ for the 
investigated temperatures. 
In this Figure the shift of the localization transition to larger $N$ 
for increasing $T$ is clearly shown. For $T> 45 \, K$ the transition 
has not been tracked down completely because the pressure required to 
reach such large $N$ values exceeds the capacity of our apparatus $(P\leq 
10.0\, MPa.)$ Nonetheless, it is evident that the localization 
phenomenon occurs also at high temperatures provided that the density 
is large enough.

\section{Discussion}\label{sec:disc}
A description of the observed behavior of $\mu_{0}N$ as a function of 
$N$ is very difficult. In fact, it must deal with the mobility of the 
two charge carriers, the extended and the localized electron, and 
also it must treat correctly the probability of occupation of the two 
states as a function of the density.

Moreover, although the mobility of the localized electron, i.e., of the bubble, is 
rather well described by the simple Stokes hydrodynamic formula, 
$\mu_{0}=e/6\pi \eta R,$ where $\eta $ is the gas viscosity and $R$ is 
the bubble radius \cite{iaku1,ec}, the description of the mobility of the extended 
electron states is still rather controversial, also because the localization 
transition is not as sharp as desired, as, for instance, in the case 
of Ne \cite{borg90}.

Several theoretical models for the description of the quasifree electron 
mobility in dense noble gases have been devised on the basis of the 
Boltzmann formalism of 
kinetic theory \cite{borg941,poli,om}. Their common 
feature is the realization that multiple scattering effects concur to 
dress the electron--atom scattering cross section. It has also been 
suggested \cite{ioffe} that, when the ratio between the 
electron thermal wavelength $\lambda_{T}$ and its mean free path 
$\ell_{c}$ is $\lambda_{T}/\ell_{c}\approx 1 ,$ the scattering rate 
diverges \cite{poli} and electrons get localized as a consequence of the 
interference of two scattering processes: the scattering off several 
different scattering centers and the time--reversed scattering 
sequence \cite{asca}. This model naturally introduces a mobility edge, an energy 
below which the electron wavefunction does not propagate. 

Although 
this mobility--edge model describes well 
the electron mobility in dense He gas, it has two main drawbacks. The 
first one is that it works correctly only for He, because its 
scattering cross section is large and nearly energy independent.
 For Ne, for instance, it does not correctly describe 
 the experimental data because of the strong dependence of the 
 momentum transfer scattering cross section \cite{borg92}. 
 
 Moreover, it is well known that, 
 in liquid He, electrons trapped in 
stable cavities within the fluid have been observed by IR 
spectroscopy \cite{adams1,adams2,hern91} and this observation has 
been confirmed also by 
quantum--mechanical Molecular Dynamics calculations \cite{kalia}, 
while the localized states, described in the mobility--edge model as
those with 
energy below the mobility edge, are only 
not propagating but do not 
reside in cavities. 
Even a static disorder produces localized 
electrons in this model. It 
is of course possible that after 
localization electrons could deform the fluid to 
produce bubble states, but the observed drop of mobility is not 
due to bubble formation \cite{poli}. 

In view of these considerations, we adopt a simple 
model \cite{miya} that describes the formation of the self--trapped
electron 
states as a process of localization in a quantum well. 
The mobility of the quasifree electrons is treated in terms of a 
different,
 heuristic model 
 developed in our laboratory that encompasses the several multiple scattering 
effects present in the scattering process of an excess electron in a 
dense gas. We use such a model because it has given excellent 
agreement with the experimental data in Ne \cite{borg88,borg90} as well as in Ar \cite{bsl}.

In addition to the usual thermal energy, electrons 
in the propagating state have a ground state energy 
$V_{0}(N)$ which depends on the density of the environment.
$V_{0}(N)$ consists of two contributions \cite{sjc}
\begin{equation}
    V_{0}(N)= E_{K}(N) +U_{P}(N)
    \label{eq:vo}
    \end{equation}
$U_{P}$ is a negative potential energy term arising from the screened 
polarization interaction of electrons with the gas atoms. $E_{K}(N)$ 
is a positive kinetic energy contribution due to excluded volume, 
quantum effects. 

\begin{figure}[htbp]
    \centering
    \includegraphics[scale=0.6]{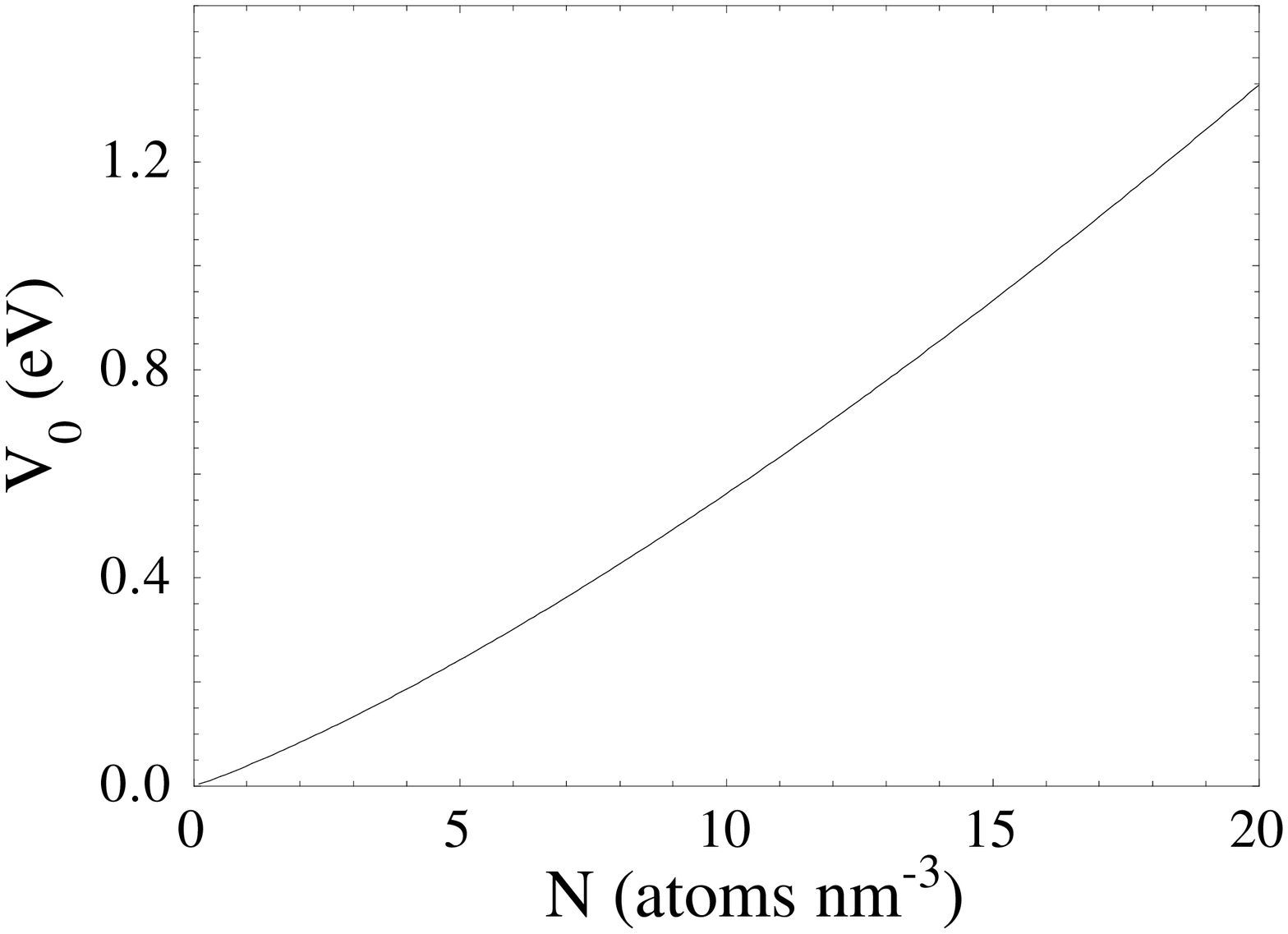}
    \caption{\small Ground state energy $  V_{0}(N)$ 
of a quasifree electron as a function of the gas density.}
    \label{fig:fig5}
\end{figure}

Owing to the small He polarizability, $U_{P}$ can be neglected thus 
yielding $V_{0}\approx E_{K}.$
It has been shown \cite{borg90,bsl,broomall} that $E_{K}$
is quite accurately
given by the Wigner--Seitz model
\begin{equation}
    V_{0}=\frac {\hbar^{2} k_{0}^{2}} {2m}, \quad \tan\left[ 
k_{0}\left(r_{s}-\tilde a\right)\right]=k_{0}r_{s}
\label{eq:eq1}
\end{equation}
as shown in Figure \ref{fig:fig5}.
In Eq. \ref{eq:eq1}, $r_{s}=(3/4\pi N)^{1/3}$ is the Wigner--Seitz radius,
$\tilde a$ is the electron-atom scattering length, and $k_{0}$ 
is the ground state
momentum of the electron. Owing to the fact 
that the electron--atom 
interaction is essentially repulsive, $V_{0}(N)$ is positive and 
increase monotonically with $N.$ 
This means that the lower is the gas density the lower is
the ground state 
energy of a quasifree electron.

Since thermally activated fluctuations 
of the density are present,
also $V_{0}$ fluctuates and electrons 
can get temporarily localized in a virtual or resonant 
state above one 
such density fluctuation where the local density is 
lower than the 
average one \cite{hern91,merz}. 

If the electron--atom interaction 
is strongly enough repulsive (as in 
the case of He) and if the fluctuation is 
sufficiently deep, there can be formation of a self--trapped 
electron state, whose stability can be determined by minimizing 
its 
free energy with respect to the quasifree state.

We therefore 
assume that localized electron resides in a quantum square well of 
spherical simmetry. The well radius is $R.$ 

Since the gas has no surface tension and since the temperature is 
pretty high for He atoms to have significant thermal energy, we must allow for some He atoms penetrating into the 
cavity and dynamically interchanging  with outside atoms.
We thus assume the bubble to be partially filled with density $N_{i}<N$ and 
filling fraction $F=N_{i}/N.$ 
The electron is thus subjected to the following spherically 
symmetric 
potential
\[ \left\{
\begin{array}{cc}
    V_{i} & \mbox{for } r<R  \\
    V_{0} & \mbox{for } r\ge R
\end{array} \right. \]
where $V_{i}$ is defined as the ground state energy of an electron 
inside the bubble.
Since $N_{i}<N,$ $V_{i}<V_{0}.$ The potential inside the bubble must 
take into account also the contribution of the polarization energy 
due to the outside gas. If the bubble were empty, the polarization 
energy ${\cal E}_{P}$ could be written as \cite{miya} 
\begin{equation}
    {\cal E}_{P}= -\frac {\alpha e^{2}}{ 2(4\pi\epsilon_{0}R)} N
    \label{eq:eq2}
\end{equation}
Since the bubble is only partially empty, to first order the polarization 
energy contribution can be written as \cite{borg90}
\begin{equation}
    {\cal E}_{P}= -\frac {\alpha e^{2}}{ 2(4\pi\epsilon_{0}R)} (1-F) N
    \label{eq:eq3}
\end{equation}
In this case the potential energy of the electron inside the bubble 
can be cast in the form 
\begin{equation} 
    V_{i}=V_{F}+{\cal E}_{P}\label{eq:eq4}\end{equation} with 
    $V_{F}=V_{0}(FN),$ i.e., the $V_{0}$ value at the density of the 
    interior of the bubble.
    
A solution of the Schr\H odinger 
equation is sought for the lowest bound $s$--wave state, 
if it exists, of energy eigenvalue ${\cal E}_{1}.$ Only the first 
eigenvalue is relevant because the temperature is quite low.  
If ${\cal R}(r)$ is the ground state solution of the radial 
Schr\H odinger 
equation,  the function $f(r)=r{\cal R}(r)$ fulfills the radial 
equation
\[ 
\begin{array}{ccc}
      \left[ {\frac {{\rm d}^{2}} {{\rm d}r^{2}}} +k_{i}^{2} \right]
f(r) =0   & \mbox{for} & r < R  \\
    \left[ {\frac {{\rm d}^{2}} {{\rm d}r^{2}}} - k_{o}^{2}\right]
f(r)=0 & \mbox{for} & r\ge R
\end{array}
\]
where $ k_{i}^{2}= (2m/\hbar^{2})({\cal E}_{1}-V_{i})$ and $k_{o}^{2}=
(2m/\hbar^{2})(V_{0}-{\cal E}_{1}).$ 
By imposing the boundary conditions on the radial wavefunction
at the bubble boundary for 
$r=R,$ we obtain the eigenvalue equation
\begin{equation}
    -\tan{X} ={ \frac X {\left(H^{2}-X^{2}\right)^{1/2}}}
    \label{eq:eq5}
\end{equation}
with $X=k_{i}R$ and $H^{2}=(2m/\hbar^{2}) (V_{0}-V_{i})R^{2}.$
If $X_{1}$ is the solution of Eq. \ref{eq:eq5}, 
then the 
energy ${\cal E}_{1}$ of the $s-$wave state is
\begin{equation}
   {\cal E}_{1} = \frac {\hbar^{2}} {2m 
R^{2}} X_{1}^{2}+V_{i}
    \label{eq:eq6}
\end{equation}

The Schr\H odinger equation admits solutions if the well 
strength is such that $H^{2}\ge \pi^{2}/4.$ This translates into 
a condition on a minimum bubble radius for the existence of a solution, namely
\begin{equation} R_{0}^{2}= 
   \frac  {\hbar^{2}\pi^{2}}{
8m\left(V_{0}-V_{i}\right)} 
\label{eq:eq7}
\end{equation}
For each value $R>R_{0}$ the eigenvalue equation Eq. \ref{eq:eq5} is solved for 
$X_{1},$ and the eigenvalue ${\cal E}_{1}$ is calculated from Eq. 
\ref{eq:eq6} 
as a function of the gas density and of the filling fraction of the 
bubble.

\begin{figure}[htbp]
    \centering
    \includegraphics[scale=0.5]{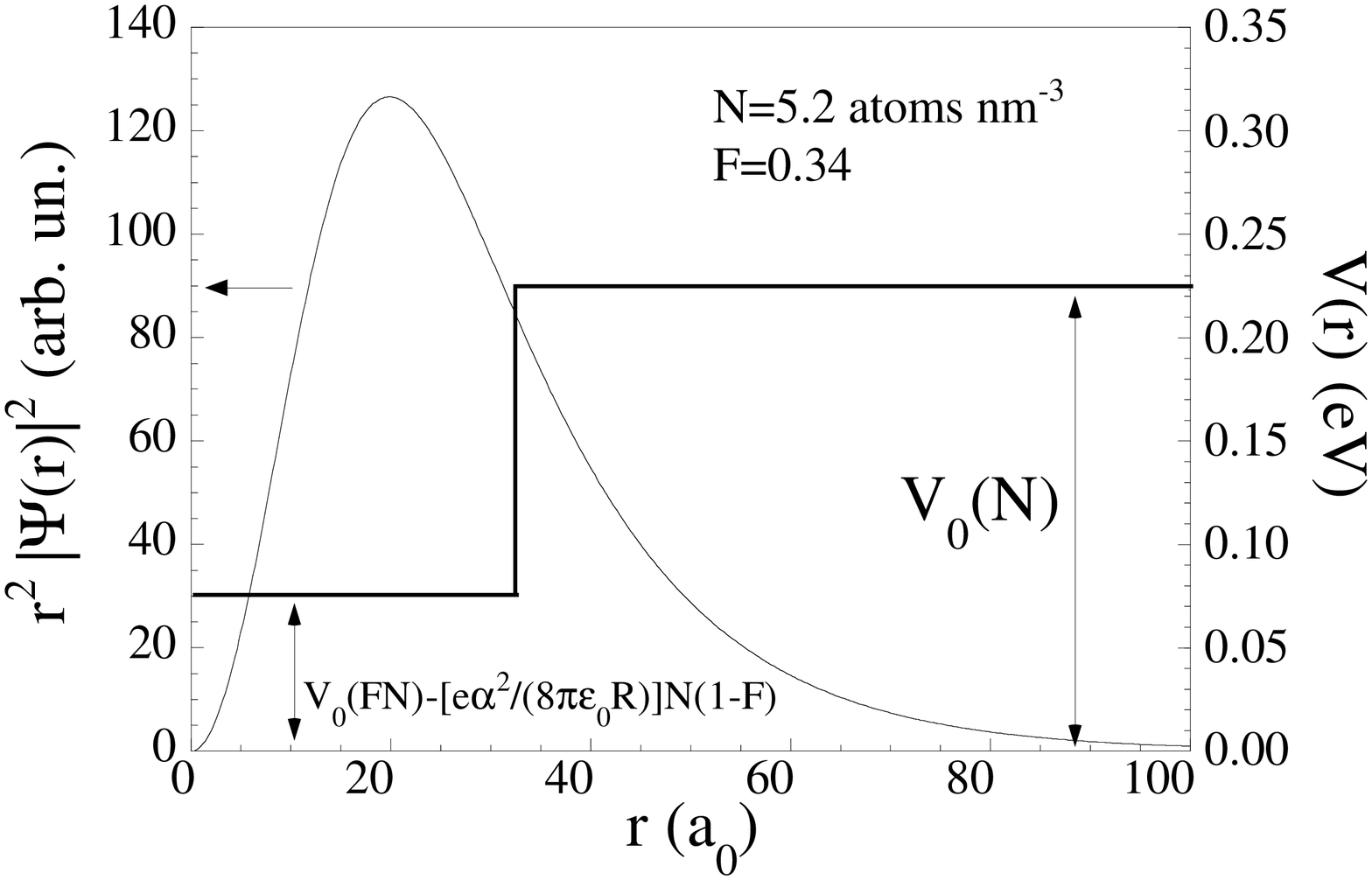}
    \caption{\small Radial probability density for the $s-$wave 
    ground state in the 
partially empty spherically symmetric square well.}
    \label{fig:fig6}
\end{figure}

In Figure \ref{fig:fig6} we show the shape of a typical $s-$ wave solution of the 
Schr\H odinger equation.
The excess free energy of the localized state with respect to the 
 delocalized one can be computed as
\begin{equation}
    \Delta {\cal F} = {\cal E}_{1} +V_{i} + W - V_{0}
    \label{eq:eq8}
\end{equation}
where $W$ is the volume work, at constant $T,$ required to expand 
 the bubble and is given by \cite{borg90}
 \begin{equation}
     W= {\frac {4\pi} 3 } R^{3}P\Bigl[ 1-F -{FN\over 
 P}\int\limits_{FN}^{N}{P\over n^{2}}{\rm d}n\Bigr]
     \label{eq:eq9}
 \end{equation}
$P$ is the gas pressure.
 In order to find the most probable state, $\Delta {\cal F}$ is 
 minimized with respect to the bubble radius and filling fraction.  
 Rigorously speaking, the minimum excess free energy should be 
 obtained by averaging $\Delta {\cal F}$ over all atomic 
 configurations leading to trapped electron states. This is a 
 formidable task and therefore, to a first approximation, we adopt 
 the optimum atom--concentration fluctuation \cite{iaku1}, i.e.,  that 
 which causes the largest decrease of the system free energy as a 
 consequence of electron trapping.
 
\begin{figure}[htbp]
      \centering
     \includegraphics[scale=0.5]{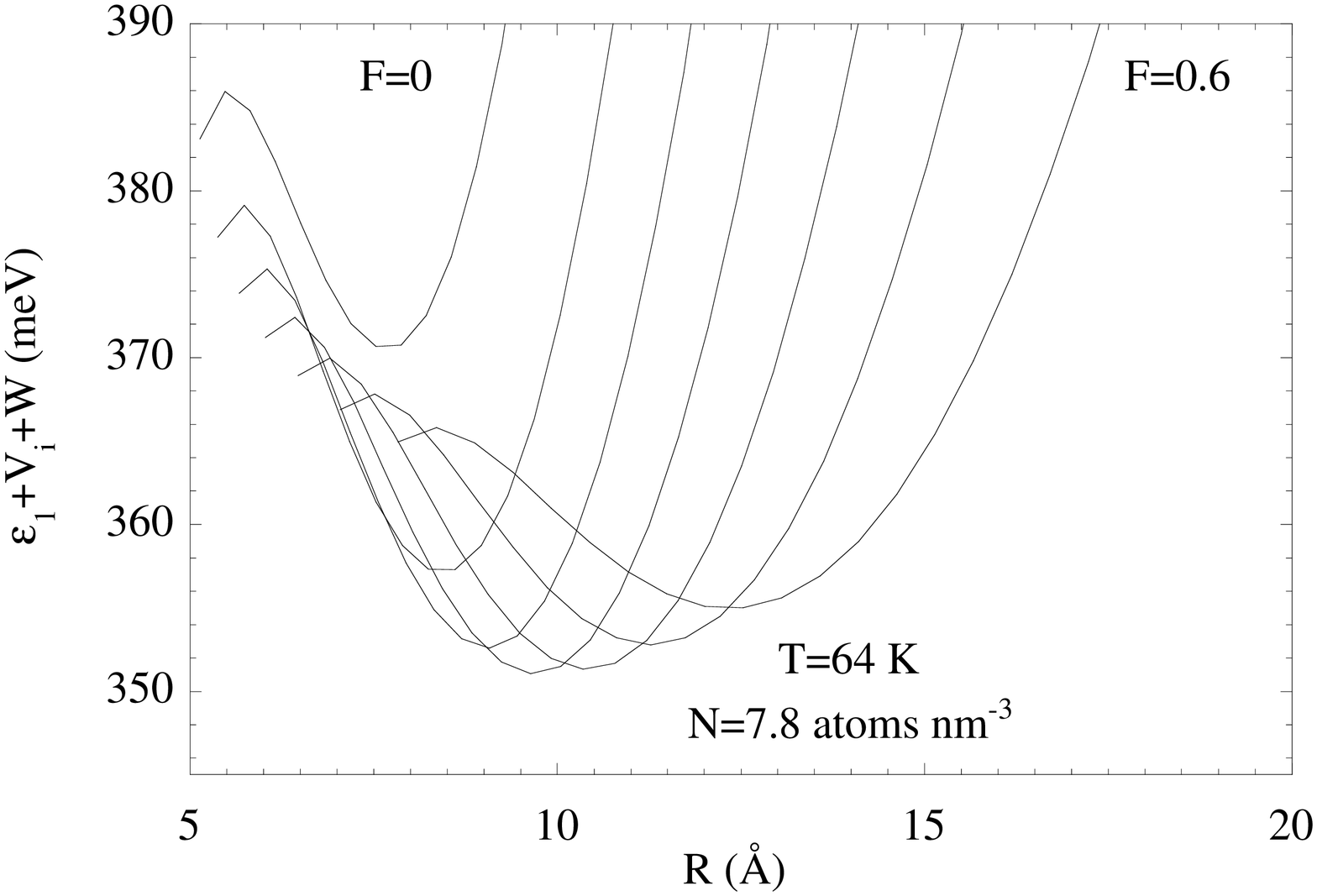}
     \caption{\small Free energy of the localized state 
 for given $  T=64
 \, K$ and $    N= 7.8\, atoms\cdot nm^{-3} $ for filling fraction $ 
    F=0,\, 0.1,\, 0.2,\, 0.3,\, 0.4,\, 0.5,\, 0.6$ as a function of the 
  bubble radius.}
     \label{fig:fig7}
 \end{figure}

\begin{figure}[htbp]
    \centering
    \includegraphics[scale=0.6]{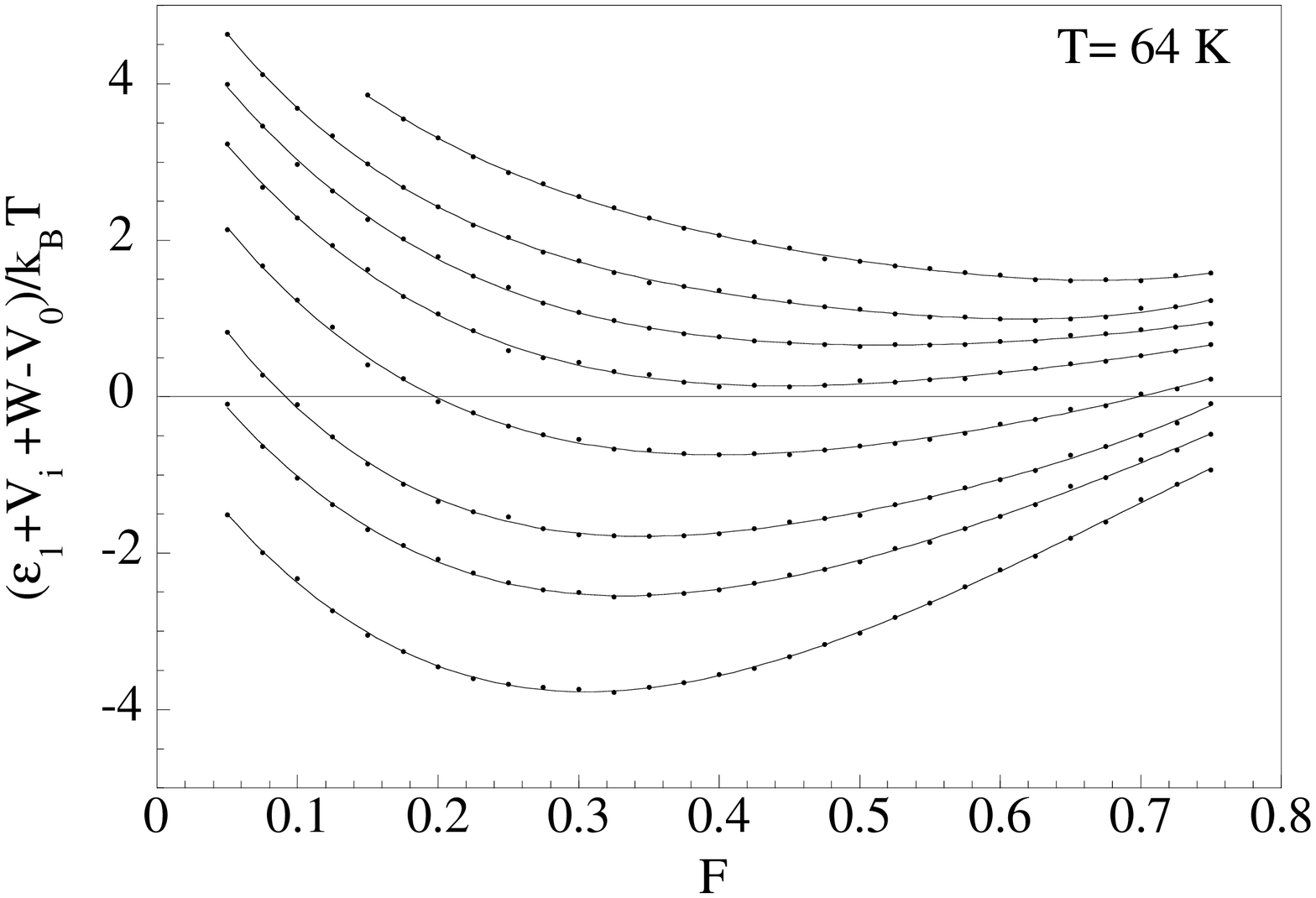}
    \caption{\small The excess free 
 energy, minimized with respect to the bubble 
radius, as a function of the filling fraction for $  5,\, 
5.3,\, 5.5,\, 5.7,\,6,\,6.3,\, 6.5,\, 6.8\, atoms \cdot nm^{-3}.$ 
for $   T = 64.4 \, K$ (from top).}
    \label{fig:fig8}
\end{figure}

  In Figure \ref{fig:fig7} we show the free energy of the localized state ${\cal 
 E}_{1}+V_{i}+ W$ as a function of the bubble radius at fixed $T=64 
 \, K$ and $N= 7.8\, atoms\cdot nm^{-3} $ for several filling fraction 
 values.  
 Once minimized as a function of the bubble radius, this first minimum 
 value of excess free energy is plotted 
in Figure \ref{fig:fig8} as a function of the filling fraction for several 
$N$ at fixed temperature.

For smaller densities, this excess free energy minimized 
with respect to bubble radius at constant $N$ and $T$ is a monotonically 
decreasing function of the filling fraction $F.$ This means that the 
incipient bubble is not stable. It gets more and more filled until it 
disappears completely. 

Stable states are now sought by carrying out a second 
minimization of excess free energy as a function of the filling 
fraction $F.$

\begin{figure}[htbp]
    \centering
    \includegraphics[scale=0.6]{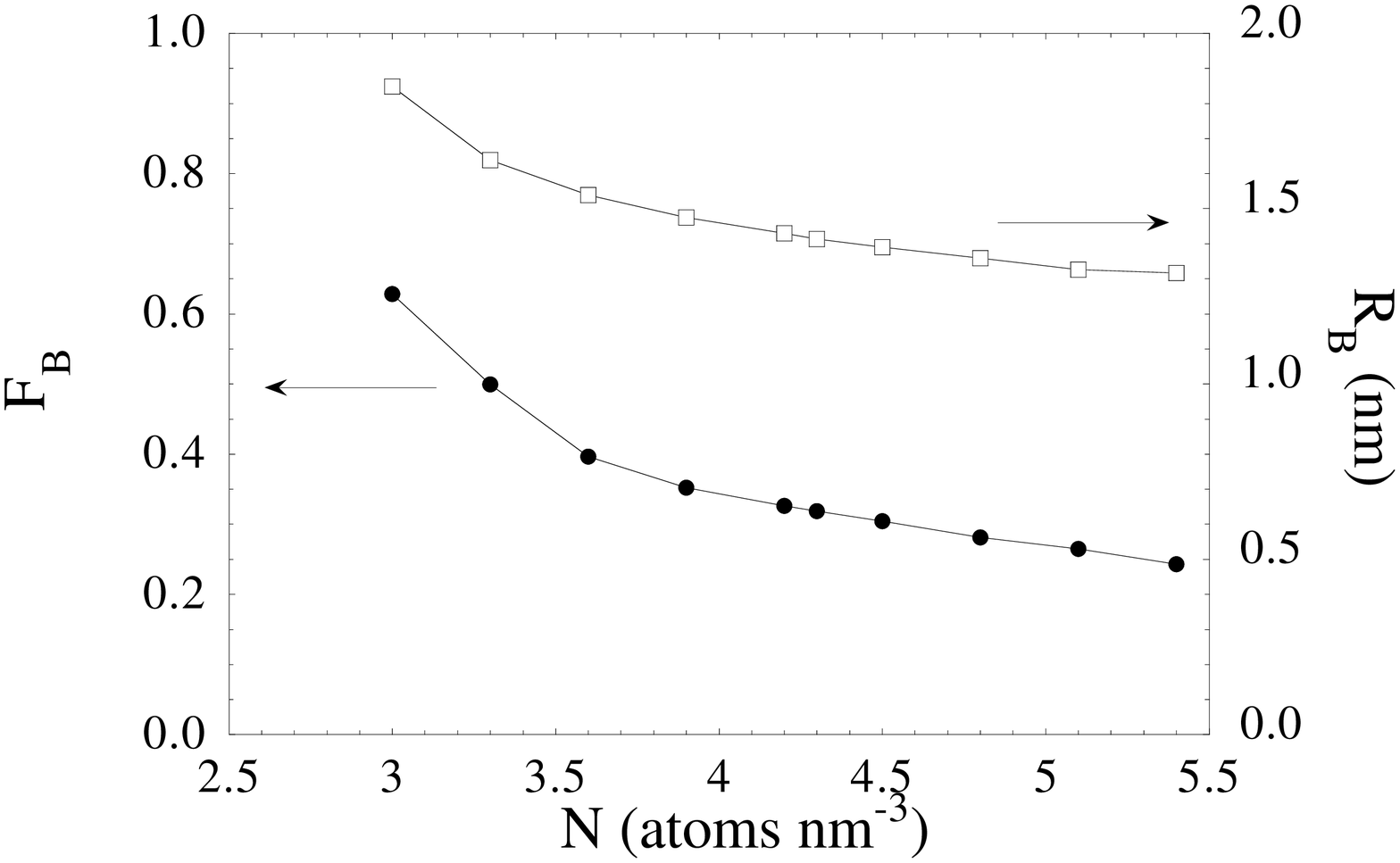}
    \caption{\small Optimum equilibrium filling fraction 
  $  F_{B}$ and radius $  R_{B}$ of the electron bubble for $
    T=26 
  \, K$ as a function of the density.}
    \label{fig:fig9}
\end{figure}

This double minimization procedure finally yields the optimum values 
of filling fraction $F_{B}$ and bubble radius $R_{B},$ shown in Figure 
\ref{fig:fig9} for $T=26 \, K$ as a function of the gas density.

\begin{figure}[htbp]
    \centering
    \includegraphics[scale=0.6]{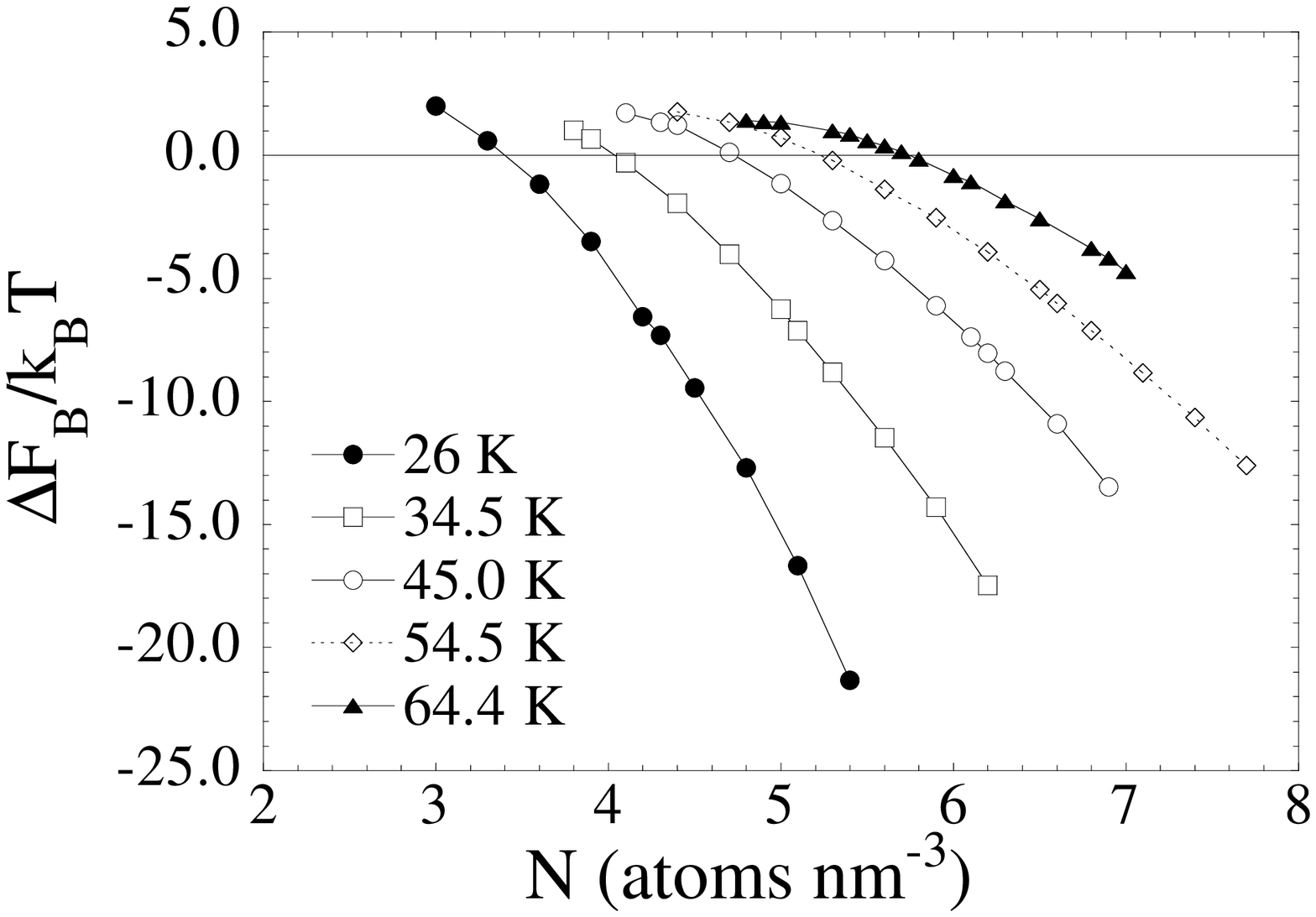}
    \caption{\small Minimum excess free energy of the localized 
  state as a function of $  N$ for several $T.$}
    \label{fig:fig10}
\end{figure}

\noindent It can be seen that, at constant $T,$ bubbles tend to become smaller and 
emptier as the density increases. The optimum bubble radius is $R_{B}\approx 1.5-2 \,\,
nm,$  compatible with 
the observed values in liquid He \cite{hern91}. 
 The values of the excess free energy corresponding to the optimum 
filling fraction and bubble, $\Delta F_{B}=\Delta F(R_{B},F_{B},N,T) ,$  
are reported in Figure \ref{fig:fig10}.

Bubble states start forming as 
soon as $\Delta {\cal F}_{B}=0,$ but they are not stable against 
thermal fluctuations 
until $\vert \Delta {\cal F}_{B}/k_{{B}}T\vert \gg 1.$ For a given $T,$ 
this condition is fulfilled only if $N$ is large enough.
Moreover, by inspecting 
Fig. \ref{fig:fig10}, we see that a given value of 
$\Delta{\cal F}_{B}$ is obtained at 
increasingly higher densities as the temperature is 
increased.  
\begin{figure}[htbp]
    \centering
    \includegraphics[scale=0.6]{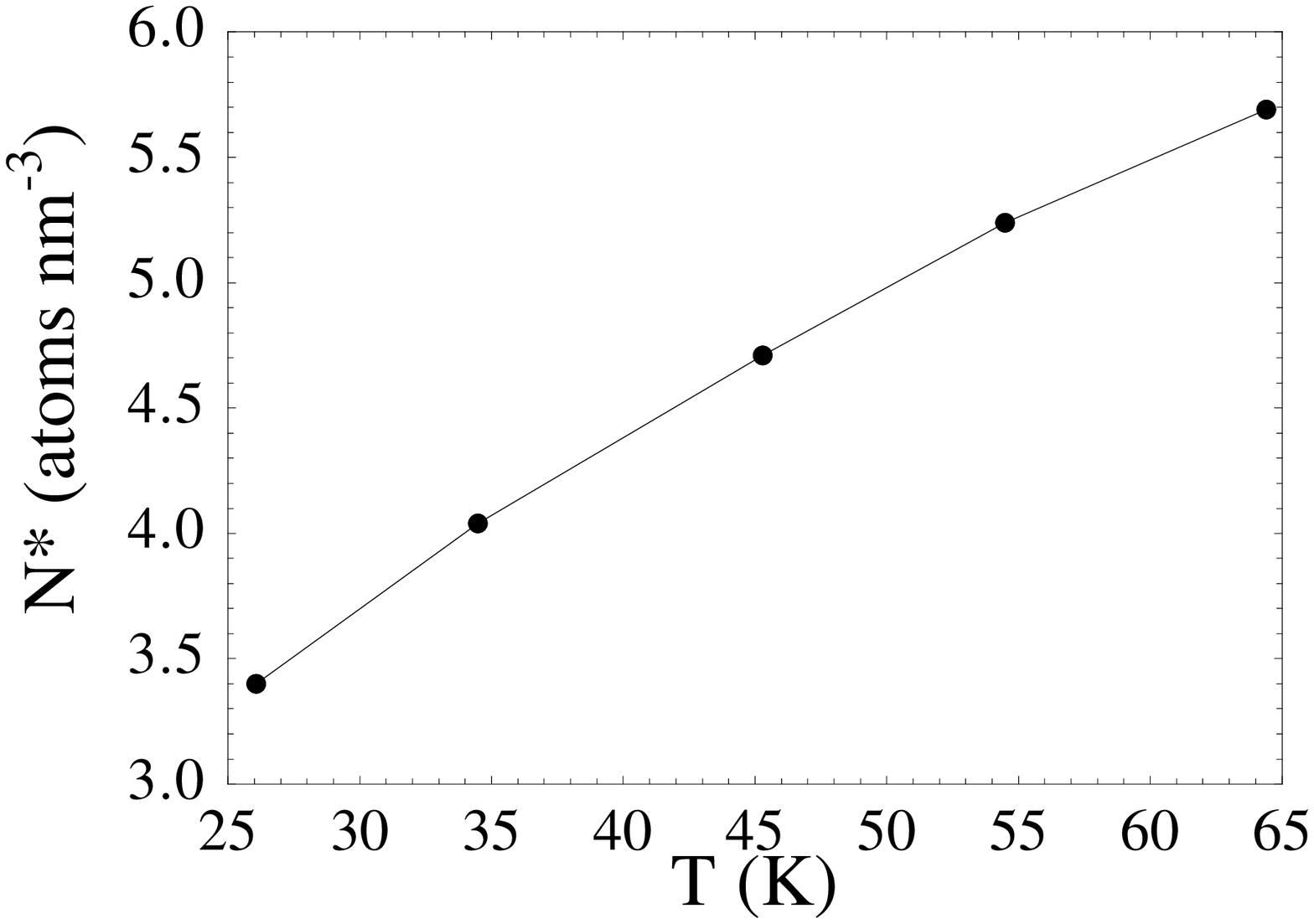}
    \caption{\small Density values $  N^{\star}$ where
   localized and delocalized states are equiprobable as a function 
  of $T.$}
    \label{fig:fig11}
\end{figure}
In Figure 
\ref{fig:fig11} we show the values of density
$N^{\star}$ where $\Delta {\cal F}_{B}=0.$ At this density the 
localized and delocalized states are equiprobable. In agreement with 
the experimental observation on mobility, $N^{\star}$ increases 
with $T.$ This means that bubbles become stable at larger $N$ when 
$T$ increases, both because electrons have more thermal energy and 
because the volume work to expand the bubble increases with the 
temperature.

Once the minimum excess free energy has been computed, the 
 fraction of bubble and quasifree states is readily calculated as 
 $n_{B}/n_{F}= \exp{\{ -\Delta {\cal F}_{B}\}}.$ The observed mobility is 
 then a weighted sum of the contribution of the mobilities of the two 
 states \cite{iaku1,young}. For the bubble state the semihydrodynamic mobility
\begin{equation}
    \mu_{B}=\frac e {6\pi\eta R_{B}}
 \left[ 1+ \frac {9\pi\eta} {4NR_{B}(2\pi m 
 k_{{B}}T)^{1/2}}\right]
    \label{eq:eq10}\end{equation}
 has been used \cite{ec}, where $\eta$ is the gas viscosity 
 \cite{wasserman}.

 For the mobility of the quasifree states we have used the results of 
 the heuristic Padua model, succesfully exploited in Ne 
 \cite{borg88,borg90} and Ar 
 \cite{bsl}. The 
 quasifree electron mobility can be written as \cite{hernmart} 
 \begin{equation}
     \mu_{F} = \frac {4e} {3hS(0)} \lambda_{T}\lambda^{\star}
     \exp{\left( 
 -\lambda_{T}/\sqrt{\pi}\lambda^{\star}\right)}
     \label{eq:eq11}
     \end{equation}
where $h$ is the Planck's constant. $S(0)= Nk_{B}T\chi_{T}$ is the long--wavelength 
limit of the static structure factor and $\chi_{T}$ is the gas 
isothermal compressibility. $\lambda_{T}=h/\sqrt{2\pi mk_{B}T}$ is 
the thermal wavelength of the electron. Finally, $\lambda^{\star}$ is 
defined as
 \begin{equation}
     \lambda^{\star} = \frac {1} {N}
     \left(k_{B}T\right)^{2}\int\limits_{0}^{\infty} 
     \frac {\epsilon} {\sigma_{mt}\left(\epsilon +E_{k}\right)} 
     {\rm e}^{\left(-\epsilon / 
k_{B}T\right)}{\rm d}\epsilon
     \label{eq:eq12}
     \end{equation}
where $\sigma_{mt}(\epsilon +E_{k})$ is the momentum transfer 
scattering cross section evaluated at the electron energy shifted by the 
kinetic contribution $E_{k}$ of the ground 
state energy shift $V_{0}.$ We recall here that, for He, $E_{k}\approx V_{0}.$  
The exponential factor in Equation \ref{eq:eq11} is due to O'Malley 
\cite{om}. 
This model includes the three main effects of multiple scattering 
\cite{bsl}:
\begin{itemize}
    \item  [1] the shift $V_{0}$ of the ground state energy of a quasifree 
electron in a medium of density $N;$

    \item  [2]  the correlation among scatterers taken into account by the 
static structure factor $S(0)$ \cite{lekner};

    \item  [3] the increase of the scattering rate due to quantum 
self--interference of electron multiply scattered in a time reversed 
sequence by the same scattering centers \cite{asca} and described by the O'Malley 
factor in Eq. \ref{eq:eq11}.
\end{itemize}
\begin{figure}[htbp]
    \centering
    \includegraphics[scale=0.6]{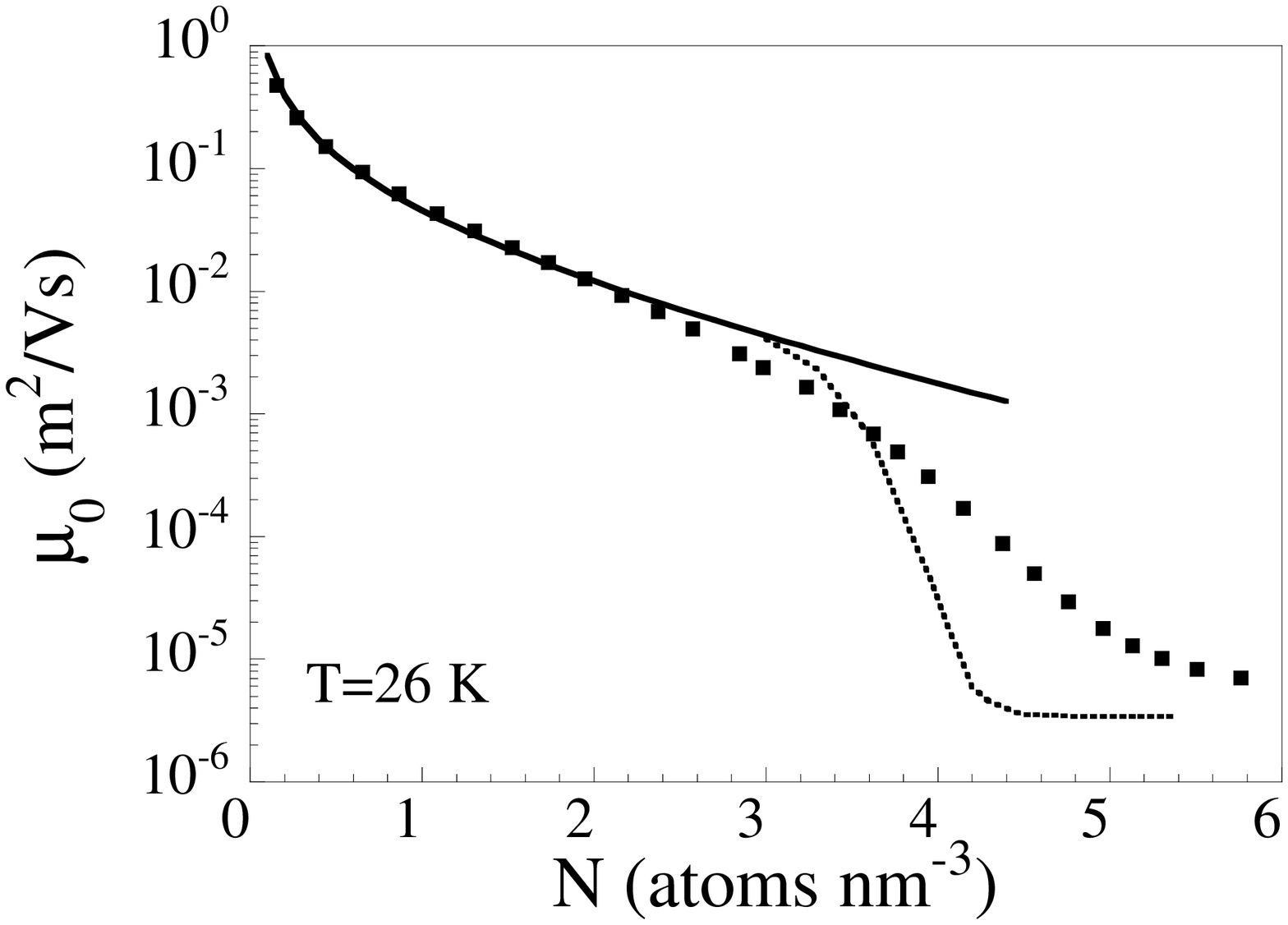}
    \caption{\small  Zero--field mobility 
 $  \mu_{0}$ vs. $  N$ for 
  $  T=26 \, {K}.$ The solid line is the mobility of quasifree 
  electron states. 
  The dashed line is the 
  weighted average mobility.}
    \label{fig:fig12}
\end{figure}
In Figure \ref{fig:fig12} we show the results of the model for $T=26\, K.$ The 
quasifree mobility in the low--density side is well described by the 
heuristic model and also the density
where the localization transition occurs is reproduced 
with satisfactory accuracy. Similar results are obtained for the 
higher temperatures. 

\begin{figure}[htbp]
    \centering
    \includegraphics[scale=0.5]{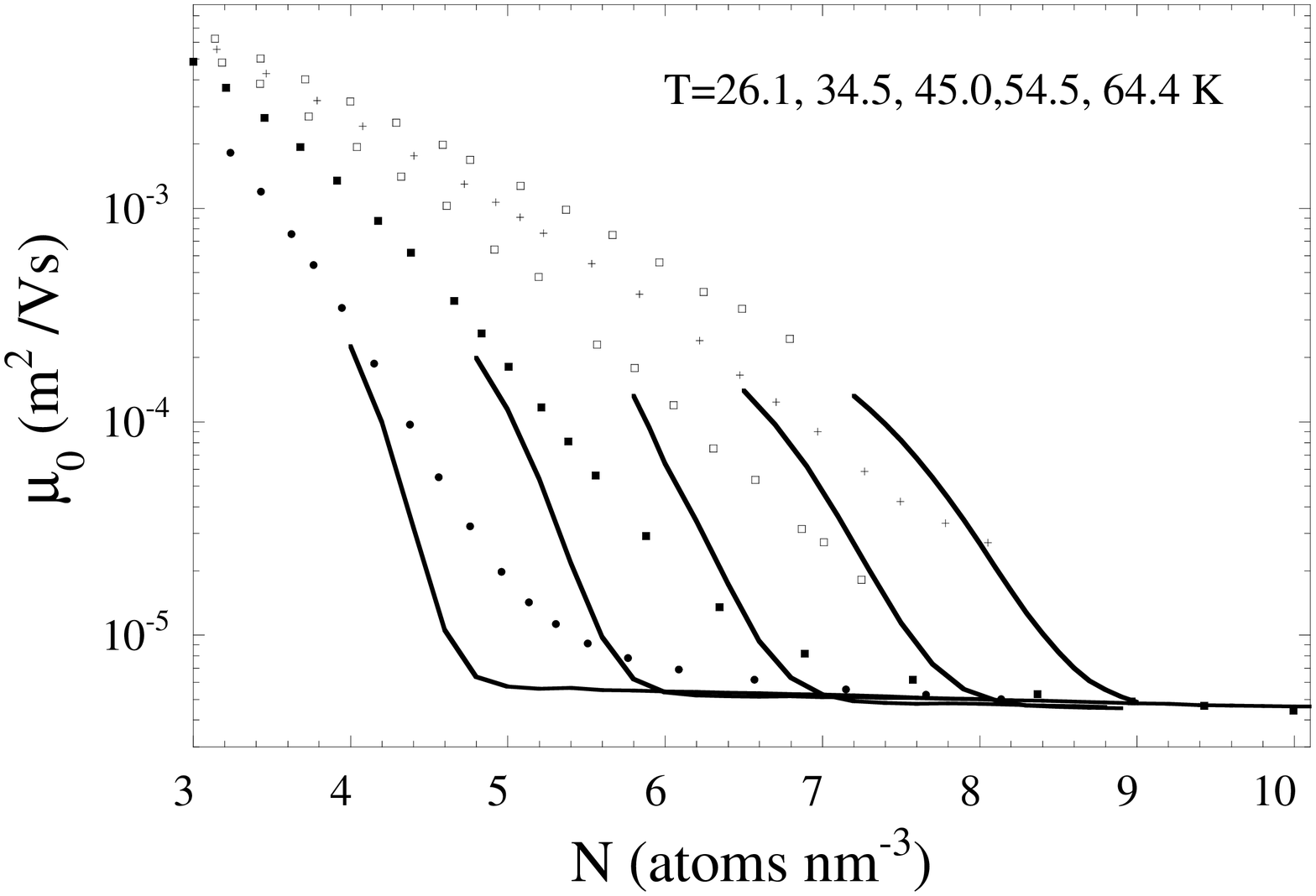}
    \caption{\small   Zero--field mobility $  \mu_{0}$ vs. $ N$ for 
  $  T=26.1,\, 34.5,\, 45.0,\, 54.5,\, 64.4 \, {K}$ in the 
  high--density region. 
  The solid line is the calculated average mobility.}
    \label{fig:fig13}
\end{figure}

In Figure \ref{fig:fig13} we show the experimental mobility in the high--density 
region for $26<T<64\, K$ with the average mobility at high density 
calculated according to the model.
This Figure clearly shows that the present model quite accurately 
predicts the shift of the localization transition to higher densities 
when the temperature is increased, although it does not fit the data 
with great accuracy.

 \section{Conclusions}\label{sec:conc}
The electron mobility in dense He gas shows two distinct regimes at low 
and high $N.$ At low $N$ the states of the excess electrons are 
extended, while at high $N$ electrons are localized in bubbles. 
Both states are present at all $N,$ but bubble states become stable, 
at fixed $T,$ only if $N$ exceeds a certain value $N^{\star}.$  The measured 
mobility is a weighted sum of the contribution of the two kind of 
electrons, quasifree and localized. 

A simple model of electron localization in a quantum square well 
explains the observed fact that the localization transition shifts to  
higher $N$ as $T$ increases. It also semiquantitatively describes the 
observed mobility. The agreement of the model with the data, however, 
is far from satisfactory. More sophisticated models, namely those 
based on the 
so--called self--consistent--field approximation \cite{iaku1,hernmart2}, 
where the density 
profile of the bubble is self--consistently calculated along with the 
electron wavefunction, can be used with results not very different 
from the present ones.

Among possible reasons to explain the 
discrepancy of the present model with the experimental data, 
there could be the fact that the bubble 
model is a simple two--state model and neglects the possibility 
that bubbles have a distribution of radii and filling fractions. 
Moreover, even the description of mobility of the quasifree electrons is 
not yet completely satisfactory.

¥


\begin{thebibliography}{99}
    \bibitem{hern91}  J.P.Hernandez, Rev. Mod. Phys., {\bf 63}, 675 
    (1991).

    \bibitem{hux74}  L.G.Huxley and R.W.Crompton, {\sl The Diffusion and Drift of 
Electrons in Gases} (Wiley, New York, 1974).

    \bibitem{borg941}  A.F.Borghesani and M.Santini, in 
    {\sl Linking the Gaseous and Condensed Phases of Matter. 
The Behavior of Slow Electrons}, L.G.Christophorou, E.Illenberger, and 
W.F. Schmidt Editors, NATO ASI Series, Vol. {\bf B 326} (Plenum, New 
York, 1994), pp. 259--279.

    \bibitem{borg942}  A.F.Borghesani and M.Santini, in 
    {\sl Linking the Gaseous and Condensed Phases of Matter. 
The Behavior of Slow Electrons}, L.G.Christophorou, E.Illenberger, and 
W.F. Schmidt Editors, NATO ASI Series, Vol. {\bf B 326} (Plenum, New 
York, 1994), pp. 281--301.
    \bibitem{levi}  J.L.Levine and T.M.Sanders, Phys. Rev. {\bf 154}, 
    138 (1967).
\bibitem{harri} H.R.Harrison, L.M.Sander, and B.E.Springett, Phys. 
Rev. {\bf B 6}, 908 (1973).

    \bibitem{sch80}  K.W.Schwarz, Phys. Rev. {\bf B 21}, 5125 (1980).
    \bibitem{ja}J.A.Jahnke, M.Silver, and J.P.Hernandez, Phys. Rev. 
    {\bf B 12}, 3420 (1975).  
    \bibitem{poli}  A.Ya.Polishuk, Physica {\bf 124 C}, 91 (1984).
    \bibitem{om}  T.F.O'Malley, J. Phys. {\bf B 13}, 1491 (1980).
   \bibitem{borg88} A.F.Borghesani, L.Bruschi, M.Santini, and G.Torzo, 
   Phys. Rev. {\bf A 37}, 4828 (1988).
 
    \bibitem{borg90}  A.F.Borghesani and M.Santini, Phys. Rev. {\bf A 
    42}, 7377 (1990).
    \bibitem{borg93}30. A.F.Borghesani, D.Neri, and M.Santini, 
    Phys. Rev., {\bf E 48}, 1379 (1993).
    \bibitem{iaku1} A.G.Khrapak and I.T. Yakubov, Sov. Phys. Usp. 
    {\bf 22}, 703 (1979).
    \bibitem{bruschi} L.Bruschi, G.Mazzi, and M.Santini, Phys. Rev. 
    Lett. {\bf 28}, 1504 (1972).   
   \bibitem{ancilotto} F. Ancilotto and F.Toigo, Phys. Rev. {\bf A 
   45}, 4015 (1992).
    \bibitem{rose1} M.Rosenblit and J.Jortner, Phys. Rev. Lett. {\bf 
    75}, 4079 (1995).
    \bibitem{rose2} M.Rosenblit and J.Jortner, J. Phys. Chem. {\bf 
    101}, 751 (1997).
    \bibitem{sakai} Y.Sakai, A.G.Khrapak, and W.F.Schmidt, Chem. Phys. 
    {\bf 164}, 139 (1992).
    \bibitem{adams1} C.C.Grimes and G.Adams, Phys. Rev. {\bf B 41}, 
    6366 (1990)
    \bibitem{adams2} C.C.Grimes and G.Adams, Phys. Rev. {\bf B 45}, 
    2305 (1992)
    \bibitem{kalia} A.Nakano, P.Vashishta, R.K.Kalia, Phys. Rev. {\bf 
    B 43}, 10928 (1991).
   \bibitem{ec}   T.P.Eggarter and M.H.Cohen, Phys. Rev. Lett. {\bf 
    27}, 129 (1971).
    \bibitem{ioffe} A.F.Ioffe and A.R.Regel, Prog. Semicond. {\bf 4}, 
    237 (1960).
    \bibitem{asca} G.Ascarelli, Phys. Rev. {\bf B 33}, 5825 (1986).
     \bibitem{borg92} A.F.Borghesani and M.Santini, Phys. Rev. {\bf A 
     45}, 8803 (1992).
    \bibitem{miya}   T.Miyakawa and D.L.Dexter, Phys.Rev. {\bf 184}, 
    166 (1969).

    \bibitem{bsl}  A.F.Borghesani, M.Santini, and P.Lamp, Phys. 
    Rev. {\bf A 46}, 7902 (1992).
    \bibitem{sjc} B.E.Springett, J.Jortner, and M.H.Cohen, J. Chem. 
    Phys. {\bf 48}, 2720 (1968).
     \bibitem{broomall} J.R.broomall, W.D.Johnson, and D.G.Onn, Phys. 
     Rev. {\bf B 14}, 2819 (1976).
     \bibitem{merz} E.Merzbacher, {\sl Quantum Mechanics} (Wiley, New 
     York, 1970).
     \bibitem{young} R.A.Young, Phys. Rev. {\bf A 2}, 1983 (1970).
     \bibitem{wasserman} V.V.Sychev, A.A.Vasserman, A.D.Kolzov, 
     G.A.Spiridonov, and V.A.Tsymarny, {\sl Thermodynamic Properties 
     of Helium} (Springer, Berlin, 1987). 
     \bibitem{hernmart} J.P.Hernandez and L.W.Martin, Phys. Rev. {\bf 
     A 43}, 4568 (1991).
     \bibitem{lekner}J.Lekner, Philos. Mag {\bf 18}, 1281 (1972).
     \bibitem{hernmart2} J.P.hernandez and L.W.Martin, J. 
     Phys.:Condensed Matter {\bf 4}, L1 (1992).
%
%
%
%
%
%
\end{thebibliography}

\end{document}